\documentclass[12pt]{article}

\usepackage{latexsym}

\font\tenmsbm=msbm10 scaled 1200
\font\sevenmsbm=msbm9
\newfam\msbmfam
\textfont\msbmfam=\tenmsbm \scriptfont\msbmfam=\sevenmsbm

\makeatletter
\@addtoreset{equation}{section}
\makeatother

\newcommand{\eref}[1]{(\ref{#1})}

\def\be{\begin{equation}}
\def\ee{\end{equation}}
\def\ba{\begin{eqnarray}}
\def\ea{\end{eqnarray}}
\def\bet{\begin{tabular}}
\def\eet{\end{tabular}}

\def\nin{\noindent}

\usepackage[dvips]{graphicx}
\usepackage{amsmath}
\usepackage{amssymb}

\catcode`@=11

\begin{document}

\begin{titlepage}

\vspace{2.5truecm}

\begin{center}

{\Large \bf  Spin-charge gauge approach to metal-insulator
crossover and transport properties in High-T$_c$ cuprates }\\

\vspace{2.0cm}

P A Marchetti$^1$, Z B Su$^2$, and L Yu$^{2,3}$\\

 \vspace{2.0cm}
 $^1$Dipartimento di Fisica ``G. Galilei', INFN, U. of
Padova, I-35131 Padova, Italy\\
$^2$ Institute of Theoretical Physics, Chinese Academy of
Sciences, 100080 Beijing, China\\
$^3$Institute of Physics, Chinese Academy of Sciences, 100080
Beijing, China

\vspace{2.5cm}

\begin{abstract}
The spin-charge gauge approach to the metal-insulator crossover
(MIC) and other anomalous transport properties in high-T$_c$
cuprates is briefly reviewed. A $U(1)$ field gauging the global
charge symmetry and an $SU(2)$  field gauging the global
spin-rotational symmetry are introduced to study the
two-dimensional $t-J$ model in the limit $t\gg J$. The MIC, as a
clue to the understanding of the `pseudogap' (PG) phase, is
attributed to the competition between the short-range
antiferromagnetic order and dissipative motion of charge carriers
coupled to the slave-particle gauge field. The composite particle
formed by binding the charge carrier (holon) and spin excitation
(spinon) via the slave particle gauge field exhibits a number of
peculiar properties, and the calculated results are in good
agreement with experimental data for both PG and `strange metal'
phases. Connections to other gauge field approaches in studying
the strong correlation problem are also briefly outlined.

\vspace{0.5cm}

\end{abstract}

\end{center}

\nin PACS numbers: 71.10.Pm, 71.10.Hf, 11.15.-q, 71.27.+a\\
\end{titlepage}

\newpage

\baselineskip 8 mm
\section{Introduction}

\noindent        The discovery of high temperature
superconductivity 20 years ago (Bednorz and M\"uller 1986) has
posed a great challenge to the
 condensed matter physics community.  It is true that there is
no consensus yet on theoretical interpretation  of  the normal state
properties and superconducting mechanism for these cuprates.
However, enormous progress has been made in material sample
preparations, experimental probes of various physical properties
using the finest tools and theoretical studies from many different
points of view. Although  physicists working in this field  still
have different opinions, most researchers would more or less accept
the following scenario:

     These cuprates share a layered structure incorporating one or more
copper-oxygen planes. The parent compounds ({\it e.g.}
La$_2$CuO$_4$) contain one electron per site which should be
metals according to the band theory, but are indeed Mott
insulators due to strong Coulomb repulsion. Upon doping beyond a
certain concentration (a few per cent) these compounds become
superconducting. Meanwhile, in the temperature range above the
superconducting transition (usually called pseudogap (PG) `phase'
on the underdoped side and strange metal (SM) `phase' on the
optimal, or slightly overdoped, side), various physical
properties, including optical, magnetic and transport, are
anomalous and cannot be described by the standard Landau Fermi
liquid (FL) theory. Most people believe that these unusual
properties are due to correlation effects  in doped Mott
insulators. One has to first study these properties using simple
models, like the Hubbard model, or $t-J$ model in the strong
coupling limit, to see whether the new physics in the PG and SM
phases beyond the FL theory can be essentially understood. Surely,
there are other effects, like inhomogeneities and electron-phonon
interactions, but most likely they are not the dominating factors
contributing to the new physics.

     The strong-correlation approach was pioneered by Anderson
(1987),  reviving his earlier work on frustrated spin 1/2  system at
triangular lattices (Anderson 1973, Fazekas and  Anderson 1874).
Instead of antiferromagnetic (AF) long range order (LRO), a spin
liquid ground state consisting of singlet resonant valence bonds
(RVB) was considered. Using the Gutzwiller-projected BCS-wave
function Anderson argued that this RVB state is naturally evolving
into superconducting state upon doping. Later, the spin-charge
separation concept (Baskaran {\it et al} 1987, Kivelson {\it et al}
1987, Zou and Anderson 1988) was introduced, with a spinon carrying
the spin degree of freedom and a holon carrying the charge degree of
freedom. Using the slave boson mean field theory (MFT) (Kotliar and
Liu 1988) and Gutzwiller approximation (Zhang {\it et al} 1988) a
d-wave superconductivity and pseudogap state were predicted using
the RVB approach. The gauge field fluctuations implementing the
single-occupancy constraints beyond the slave boson MFT were
considered by a number of authors (Baskaran and Anderson 1988, Ioffe
and Larkin 1989, Nagaosa and Lee 1990). Very recently, a
comprehensive review on the strong-correlation approach to high
$T_c$ (HTS) cuprates, focusing on the gauge theory approach was
provided by Lee {\it et al} (2006).

    For the last few years we have developed a different gauge
approach to attack the HTS problem, using a $U(1)$ field to gauge
the global charge symmetry and an $SU(2)$ field to gauge the
global spin symmetry ( Marchetti {\it et al} 1998, Marchetti {\it
et al} 2000, Marchetti {\it et al} 2001,  Marchetti {\it et al}
2004a,b, Marchetti {\it et al} 2005). Unlike the slave-boson
approach, the charge degree of freedom in our approach is carried
by a fermion, while the spin degree of freedom is represented by a
boson. We found this approach to be very efficient in treating the
most pronounced phenomenon in the PG phase, namely the
metal-insulator crossover (MIC) as well as  various transport,
optical and magnetic properties. In this short review we will
outline the basic ideas of our approach and summarize our main
results in comparison with experiments. We will attempt to make
the presentation more readable and emphasize the physical
interpretation, referring readers to original papers for technical
details.

\section{Metal-insulator crossover  as a clue  to the HTS problem}

\noindent  The PG phase was anticipated by the RVB approach and
has been
 studied very thoroughly in an enormous amount of experimental and
 theoretical work which has been very well summarized in recent reviews (Timusk
 and Statt 1999, Norman and Pepin 2003, Lee {\it et al} 2006). In
 our view the most spectacular phenomenon of the PG phase is the MIC
observed in underdoped, non-superconducting cuprates in the
absence of magnetic field and a similar phenomenon in
superconducting samples when an applied strong magnetic field
suppresses the superconductivity.

First of all, this MIC is a rather universal phenomenon. A minimum
in resistivity (around 50 - 100K) and a crossover from metallic
conductivity ${d\rho \over dT} > 0$  at high temperatures to
insulating behavior ${d\rho \over dT} < 0$ at low temperatures has
been observed in heavily underdoped LSCO (Takagi {\it et al} 1992,
Keimer {\it et al} 1992, Ando {\it et al} 2001, Ando {\it et al}
2004), non-superconducting Bi$_{2+x}$Sr$_{2-y}$CuO$_{6\pm\delta}$
(Fiory {\it et al}), non-superconducting YBCO (Wuyts {\it et al}
1996, Trappeniers {\it et al} 1999, Ando {\it et al} 1999, Ando
{\it et al} 2004) and La-doped Bi-2201 (Ono and Ando 2003).  It
has also been observed in electron-underdoped NCCO (Onose {\it et
al} 2001) and PCCO ( Fournier {\it et al} 1998).

Moreover, such a MIC has also been observed in a number of
superconducting samples when a strong magnetic field suppresses
the superconductivity, including  LSCO (Ando {\it et al} 1995,
Ando {\it et al} 1996a, Boebinger {\it et al} 1996), La-doped
Bi-2201 (Ando {\it et al} 1996a,b, Ono {\it et al} 2000),
electron-doped PCCO (Fournier {\it et al} 1998) as well as in
Zn-doped YBCO (Segawa and Ando 1999).

 This phenomenon has not had much attention so far, because many people attributed
 it to localization due to disorder. However, that interpretation is at odds with the
 following facts:
 (1) Including higher
doping samples exhibiting MIC in strong magnetic fields, the
estimate for $k_F\ell$, where $k_F$ is the Fermi momentum and $\ell$
the mean free path, ranges from 0.1 to 25  at the MIC, {\it i.e.}
from far below to far above the Ioffe-Regel limit ($k_F\ell \sim
1$), characterizing the MIC due to disorder localization.
 (2) In LSCO samples with $a-b$ in-plane anisotropy
the MIC temperature of $\rho_a$ is different from that of $\rho_b$
(Dumm {\it et al} 2003) contradicting the `unique' localization
temperature, characteristic of the (at least standard) theory of
localization.
 (3)A universality of suitably
normalized resistivity (Wuyts {\it et al} 1996, Konstantinovic {\it
et al} 2000) has been observed  in terms of $T/T^*$, where $T^*$ is
roughly proportional to $T$ at the MIC ($T_{MIC})$ and can be
identified as the PG temperature. All the above features are very
difficult, if not impossible to explain from the localization
viewpoint. On the other hand, as discussed below, the spin-charge
gauge approach provides a rather natural interpretation for these
unusual features (Marchetti {\it et al} 2004a,b).

Furthermore, once the MIC is recognized as an intrinsic property
of cuprates, it is very difficult in conventional theory to
reconcile the insulating behavior at low temperatures with the
presence of a finite Fermi surface (FS), as shown by
angle-resolved photoemission spectroscopy (ARPES) (See, {\it
e.g.}, Damascelli {\it et al} 2003). Meanwhile, these two
seemingly contradictory phenomena can be easily accommodated in a
slave-particle gauge theory, due to the Ioffe-Larkin (1989) rule,
stating that the inverse conductivity of the electron is the sum
of the inverse conductivity of the spinon and the inverse
conductivity of the holon. This non-standard feature can be
intuitively understood as a consequence of the gauge string
binding spinon to holon: the velocity of the electron is
determined by the slowest (not the fastest!) among spinon and
holon. Then, if the fermionic excitation, spinon or holon
depending on the approach, has a FS and therefore a metallic
resistivity vanishing at $T \sim 0$, the electron can still have a
metallic/insulating behavior at low $T$ if the bosonic excitation
does. If the leading contribution comes from the boson, without a
detailed reference to FS, this can qualitatively explain the
universality quoted in point (3) above.

In our view the MIC {\it is the clue to the understanding of the
PG phase}. The MIC in underdoped cuprates in the absence of
magnetic field and MIC in superconducting samples when a strong
magnetic field suppresses superconductivity {\it  is the same
phenomenon with the same origin}: as an outcome of competition
between the AF short range order (SRO) and the dissipative motion
of the charge carriers. We start from the Mott insulating state
showing AF LRO. Upon doping beyond certain threshold the AF LRO is
destroyed, being replaced by SRO, characterized by an AF
correlation length $\xi$. Since  holes distort the AF background
and their average distance $\sim \delta^{-1/2}$, intuitively, $\xi
\approx \delta^{-1/2}$, where $\delta$ is the doping
concentration. This has been confirmed by the neutron scattering
experiments (Keimer {\it et al } 1992). Our theoretical treatment
proves $\xi \approx (\delta|\ln{\delta}|)^{-1/2}$, providing the
first length scale, while the corresponding energy scale is the
spin excitation (spinon) gap $m_s=J(\delta|\ln{\delta}|)^{1/2}$,
where $J$ is the AF exchange interaction. A competing factor is
the diffusive motion of the charge carriers with characteristic
energy $\sim Tm_h$, where $T$ is the temperature, while $m_h \sim
\delta/t$ is the effective mass of the charge carrier (holon) in
the PG phase, $t$ is the hopping integral. The corresponding
length scale is the thermal de Broglie wave length $\lambda_T \sim
(T\delta/t)^{-1/2}$. At low temperatures $\xi \lesssim \lambda_T$,
the AF SRO dominates and the charge carriers become weakly
localized (not exponentially) showing insulating behavior. We
would like to emphasize that this `peculiar localization' is
mainly due to interaction rather than disorder, and it is also
different from the standard Mott insulator which comes entirely
from the Coulomb repulsion. On the contrary, at high temperatures
$\xi \gtrsim \lambda_T$, the diffusive motion of charge carriers
prevails, exhibiting metallic conductivity. Therefore the
competition of the real part of the spinon `self-energy', the mass
gap, and the imaginary part, the dissipation, gives rise to this
spectacular phenomenon: MIC.

    It turns out that a number of experimental observations in the
PG phase can also be explained by the competition of these two
factors and the `composite' nature of low-energy excitations.
Roughly speaking, the spin (spinons) and charge (holons) excitations
in the PG phase behave like `separate particles' in their scattering
against gauge fluctuations, which renormalizes their properties and
dominates the in-plane transport phenomena. However, at small
energy-momentum scale the gauge field binds spinon and anti-spinon
into magnon `resonance'. Similarly spinon and holon are bound into
electron `resonance' with non-Fermi-liquid properties, showing up in
ARPES experiments as a `quasiparticle peak' and responsible for
interlayer transport. In particular, the `composite' nature of
particles is exhibited in the anomalously large spectral weight away
from the quasi-particle peak seen in ARPES (See, {\it e.g.}, Nagaosa
and Lee 1990, Lee and Nagaosa 1992, Laughlin 1997, Orgad {\it et al}
2001). In a sense, the low-energy excitations in these strongly
correlated materials are no longer well-defined quasi-particles of
FL theory, but rather these loosely bound `composite particles',
taken as a first step going beyond the well established condensed
matter physics paradigm.

\section{Spin-charge decomposition: statistics and symmetries}

\noindent Our basic assumption is that the low-energy physics of
the Cu-O layers in HTS cuprates can be described qualitatively by
a two-dimensional $t-J$ model:
 \be \label{1} H = P_G \sum_{<ij>}\Bigl(
- t \sum_\alpha c^*_{\alpha i} c_{\alpha j} + J (\vec S_i \cdot \vec
S_j - {1\over 4} n_i n_j) \Bigr) P_G, \ee

 \nin where  $i$ correspond
to Cu--sites, $\alpha=1,2$ is the spin index and $P_G$ denotes the
Gutzwiller projection, eliminating the double occupancy.
Numerical simulations yield  $t \simeq 0,4 eV$ , $J \simeq 0.13
eV$ which we use in calculations of physical quantities.
 The qualitative behavior of the low--energy physics in the
$c$--direction is obtained by adding an interlayer hopping term.

Once the $t-J$ model is accepted, a key question is how to handle
the Gutzwiller projection. A straightforward way is either to use
numerical techniques (Gros 1988, 1989, Paramekanti {\it et al}
2001, 2003), or  MFT ( Zhang {\it et al} 1988, Anderson {\it et
al} 2004). Another way is to decompose the electron into holon and
spinon: \be \label{2} c_\alpha = h s_\alpha \ee

 A simple counting of degrees of freedom (dof) shows that
the decomposition \eref{2} has an unphysical redundancy, {\it
i.e.}, a local gauge symmetry. In fact, the electron has 4 dof
($c_\alpha, c_\alpha^*)$, while the spinon $s_\alpha$ being still
of two components, also has 4 dof, and the holon $h$ being
spinless has 2 dof. One has to impose the constraint of no double
occupancy, which in slave particle formalism becomes holonomic.
Then the redundant dof (4+2-1=5) corresponds to a local phase
factor by which the holon can be multiplied, while  the spinon is
multiplied by its inverse: a $U(1)$ gauge symmetry. It is very
useful to make manifest the `hidden' gauge symmetry by introducing
an emergent $U(1)$ gauge field, coupling spinons to holons.

\subsection{Different choices of spin-charge decomposition}

\noindent Accepting the spin-charge decomposition as a good
technique to explore the low--energy physics of cuprates, one
still has several options for its implementation. The first choice
regards the statistics of spinon and holon. If no approximations
are made all these choices are equivalent and a proof of this
equivalence for some of them is given in  Fr\"ohlich and Marchetti
(1992). However, as soon as approximations are made, like mean
field (MF), they become intrinsically different. Standard choices
are: $s_\alpha$ fermion and $h$ hard-core boson or $s_\alpha$ hard
core boson and $h$ fermion. These choices are at the basis of
traditional slave boson (Zou and Anderson 1988, Nagaosa and Lee
1990, Lee and Nagaosa 1992)  and slave fermion (Arovas and
Auerbach 1988, Yoshioka 1989, Dorey and Mavromatos 1991)
approaches. To describe PG the more sophisticated approach of
slave boson pursued in Wen and Lee (1996), Lee {\it et al} (1998)
is obtained by replacing a single boson $h$ with a pair of bosons
$h_\alpha$, enhancing the slave-particle symmetry from $U(1)$ to
$SU(2)$. A slave-fermion version of this is discussed in Farakos
and Mavromatos (1998).

 Moreover, in 2D there are also other `anyon' statistics
available, in particular the `semion statistics' advocated
originally by Laughlin (1988), in which an interchange of the
fields produces a factor $e^{\pm i {\pi\over 2}}$. The product of
two semions is still a fermion: $(e^{\pm i {\pi\over 2}})^2 = -1$.
Notice also that for semions a kind of `hard-core exclusion' holds
(Fr\"ohlich and Marchetti 1992), here ensured by $P_G$.

 To adequately describe the FS of cuprates determined  by
ARPES (Damascelli {\it et al} 2003) is an important guideline in
making choices: for optimally doped materials one finds a FS
agreeing with band calculations. This is clearly compatible with
$s_\alpha$ being a fermion, as in the slave boson approach, but it
is troublesome for the slave fermion approach because $h$ being
spinless the MF Fermi momentum is expected to be doubled  w.r.t.
to the electron case (Lee {\it et al} 2006). For a semion,
although the situation is not completely elucidated (Wu 1994,
Haldane 1994, Chen and Ng 1995), it appears that a kind of
generalized Pauli--principle holds, related to Haldane statistics
(Haldane 1991), stating that for low $T$ one can accommodate at
most 2 (spinless) semions in the same momentum state and the
semion distribution function is approximately twice the fermion
distribution function. Hence a gas of spinless semions of finite
density should give a `FS' coinciding with that of spin ${1\over
2}$ fermions of the same density. We take here this generalized
Pauli principle for granted.

\subsection{Hint from the 1D $t-J$ model}

\noindent The 1D $t-J$ model in the limit $J/t \searrow 0$ has
been solved exactly by the Bethe Ansatz method (Ogata and Shiba
1990), while the correlation functions have been calculated using
the conformal field theory (CFT). Applying the spin-charge
decomposition approach we have reproduced the main results of
these techniques within a kind of MFT (Marchetti {\it et al }
1996). A coordinate representation of the solution exhibits the
following features:

\noindent (1) Spin-charge separation: the charge and the spin dof
are characterized by different physical behavior, in particular
their velocities are different $v_s \not = v_c$ (Luttinger liquid)

\noindent (2)The low--energy physics of the charge dof is
described by a free spinless fermion (holon), while the low-energy
physics of the spin dof is described by an spin ${1\over 2}$ AF
Heisenberg model in a squeezed spin chain, obtained by omitting
the unoccupied sites of the original 1D lattice, where the spins
are given by $\vec S = S^*_\alpha {\vec \sigma_{\alpha\beta}\over
2} S_\beta$ with $S_\alpha$ a `renormalized' gapless spinon.
$S_\alpha$ can be identified as a Gutzwiller projected fermion in
the squeezed chain (Marchetti {\it et al } 1996). The scaling
limit of such model is the $O(3)$ nonlinear $\sigma$-model with
$\Theta$ vacua at $\Theta=\pi$, the value when gapless excitations
appear in 1D.

\noindent (3) The electron field is a product of a holon and a
spinon field together with a non--local 'string phase': $
c_{\alpha x} = h_x \exp[{{i\over 2} \Sigma_{j < x} h_j^* h_j}]
S_{\alpha x}. $ It can be shown (Marchetti {\it et al } 1996) that
both $S_{\alpha x}$ and \be \label{4} h_x e^{i {\pi\over 2}
\Sigma_{j < x} h^*_j h_j} \ee

\noindent describe semions. One can prove that $S_\alpha$ is just
a MF approximation of a non--local spinon field with a `string
phase' analogous to that in \eref{4}, constructed out of bosonic
spinons $s_\alpha$, and its explicit expression is given in
Marchetti {\it et al } (1996). The semion statistics is crucial
for obtaining the correct correlation functions in the scaling
limit (Marchetti {\it et al } 1996) (see also Ha and Haldane
1994).

One should note that this statistics is compatible with an
optimization of energy: in the semion approach the spinon
$s_\alpha$ appears in the hopping term in the Affleck-Marston (AM
1988) form AM=$\sum_\alpha s_{\alpha i}^* s_{\alpha j}$ while in
the AF-Heisenberg term in the RVB form, RVB=$\sum_{\alpha \beta}
\epsilon_{\alpha \beta} s_{\alpha i} s_{\beta j}$, as $|$RVB$|^2$.
This occurs because for a holon-empty site $j$ the spinon
contribution turns out to be given by
\be \label{id}
\begin{pmatrix} \tilde s_{1 j}\cr \tilde s_{2 j}\end{pmatrix} \equiv
\begin{pmatrix}s_{1 j} & - s^*_2 \cr s_{2 j}& s_{1 j}^*\end{pmatrix}
\sigma_x^j
\begin{pmatrix}1 \cr 0 \end{pmatrix},
\ee
 \noindent where the spin flip induced by $\sigma_x$ is due to an
emergent AF structure. However, for a holon-occupied site there is
an additional spin flip and this is the case for the final site of
a hopping link. The peculiar feature of this spin flip occurs only
in a $SU(2)$-formalism. The interesting point is that the
following identity holds for $s_\alpha$ hard-core boson:
$|$AM$|^2$+$|$RVB$|^2$=1. Optimization of the $t$ term suggests
$|<$AM$>|$=1 but due to the previous identity this choice
optimizes also the positive $J$ term, since it corresponds to
$|<$RVB$>|$=0.

 From \eref{4} one can see that a semion can be
constructed out of fermions (and in similar  way also from
hard-core bosons) through a generalization of the Jordan-Wigner
transformation (Fradkin 1991) with an exponent which is half of
the standard one. The continuum analogue of \eref{4} is
 \be
\label{5} h (x) e^{i {\pi\over 2} \int H (x-y) h^* (y) h(y) dy}
\ee

\noindent where $H(x)$ is the Heaviside step function. The
representation \eref{5} makes it clear that the `statistics
transmutation' (from fermion to semion) in 1D is obtained by
adding an exponential of a `step function' or `kink' average of
the density.

\subsection{Generalization to two dimensions}
\noindent If we accept the suggestion coming from the 1D model for
the statistics of holon and spinon, one should search for a
semionic representation of the electron field in 2D. A semion
field in 2D is constructed out of a fermion as in \eref{5} by
replacing the 1D `kink' $H (x-y)$ with a 2D `vortex' arg $(\vec x
) \equiv \arctan ({x_2\over x_1})$ in the average of the density.

 It is convenient to rewrite the resulting exponential in
terms of a gauge field
 \be \label{6} \vec B (y) ={1\over 2} \int
\vec\nabla \arg (\vec y - \vec z) h^*h (z, y^0) d^2 z \ee
\noindent as $\exp[{i\int_x^\infty \vec B (y) \cdot d\vec y}]$

 To show that indeed $\vec B \equiv \{B_\mu \quad \mu =
1,2\}$ is the gauge field associated with a vortex let us compute
its field strength, using $\vec\nabla \times \vec\nabla$ arg
$(x-y) = \delta (x-y)$ we get
 \be \label{7} \vec\nabla \times \vec
B (y) = {1\over 2} \rho (y), \ee

\noindent where $\rho$ is the density of $h$ (Notice that from
$\vec\nabla \cdot \vec\nabla$ arg $(\vec x-\vec y)=0$ it follows
$\vec\nabla \cdot \vec B=0$, {\it i.e.} $\vec B$ is in the Coulomb
gauge). Equation \eref{7} is the same  as that appearing in the
fractional quantum Hall effect (FQHE) (see, {\it e.g.} Fradkin
1991) replacing ${1\over 2}$ with ${1\over \nu}$ where $\nu$ is
the filling at the plateau. Hence the vortices appearing in
\eref{6} are analogous to those introduced by Laughlin in the FQHE
and in fact a semionic representation of the electron was
advocated by him (Laughlin 1988) in the early days of HTS.

 Actually the physics here is  dual to that of FQHE:
there the flux was integer and the charge fractional,  while here
it is the opposite. The above ideas have been made precise, within
the spin-charge gauge approach, in Marchetti {\it et al} (1998),
where the following decomposition of the electron has been proved,
in terms of a fermionic holon and a bosonic spinon $s_\alpha$
satisfying the Gutzwiller constraint $\sum_\alpha s^*_\alpha
s_\alpha =1 $: \be \label{8} c_x \sim h_x e^{i\int^\infty_x \vec B
(y) \cdot d\vec y} P (e^{i\int^\infty_x \vec V(y) d\vec
y})_{\alpha\beta} s_{\beta x}, \ee

\noindent where $\vec B$ and $\vec V$ are  U(1) and $SU(2)$ gauge
fields, respectively, and $P$ is a path--ordering needed for the
non--abelian nature of $V_\mu$. In \eref{8} we have \ba \label{9}
\vec B (y) = {1\over 2} \sum_j \vec\nabla \arg (\vec y - j) (1 - h_j^* h_j(y^0)),\\
\vec V (y) = \sum^3_{a=1} \sigma_a \vec\nabla\ arg (\vec y - j) (1 -
h^*_j h_j(y^0)) (\tilde s_{\alpha j} \sigma^a_{\alpha\beta} \tilde
s_{\alpha j}(y^0)) \label{10} \ea

\noindent with $\tilde s_{\alpha j} =s_{\alpha j}$ if $j$ is on
the even N\'eel sublattice and $\tilde s_{\alpha j} =
\epsilon_{\alpha\beta} s^*_{\beta j}$ if $j$ is on the odd N\'eel
sublattice (compare with \eref{id} ). Here an AF structure emerges
formally and the basic related assumption is that  fields
$s_\alpha$ have a good continuum limit, but not $\tilde s_\alpha$.
The physical spin turns out to be \be \label{11} \vec S_j \simeq
\tilde s^*_{\alpha j} \vec \sigma_{\alpha\beta} \tilde s_{\beta j}
= (-1)^{|j|} s^*_{\alpha j} \vec \sigma_{\alpha\beta} s_{\beta j},
\ee \noindent where $|j|$=0 (1) for a site in the even (odd)
N\'eel sublattice, so that the AF ordering naturally  evolves as
$\delta\searrow 0$ to the LRO  of the 2D AF Heisenberg model, in
agreement with experiments  in undoped cuprates. In \eref{8} both
$h_x e^{i\int_x^\infty \vec B, d\vec y} $ and $(P e^{i
\int_x^\infty V d\vec y})_{\alpha\beta} s_{\beta x}$ are semion
fields.

\subsection{Our choice of mean field approximation}
\noindent Where do the $U(1)$  and the $SU(2)$ groups of \eref{9},
\eref{10} came from? They are just the global charge- $U(1)$ and
spin-$SU(2)$ symmetry (not to be confused with the $SU(2)$
slave-particle symmetry of (Wen and Lee 1996, Lee {\it et al}
1998)) of the $t-J$ model. In  Marchetti {\it et al} (1998) it is
shown that if these spin-charge symmetries are made local by
introducing a coupling with a $U(1)$ field $\vec B$ and a $SU(2)$
field $\vec V$ given by \eref{9}, then the obtained `spin-charge
gauged $t-J$ model' is strictly equivalent to the original $t-J$
model \footnote{In path integral formalism this is obtained within
a Chern-Simons theory}. Of course, as soon as MF approximations
are made, this approach becomes distinct from {\it e.g.} the slave
boson or slave fermion,  becoming a new `slave-semion' approach. A
variant of this is obtained by replacing $SU(2$) of spin with the
$U(1)$ subgroup unbroken in the AF phase. This is the choice made
by Weng and collaborators (Weng {\it et al} 2000, Weng 2003).

 However, in the `semionic' approach it is too hard to
keep charge, spin and slave particle symmetries all exact and an
approximation has been made to perform explicit computations. In
the spin-charge gauge approach pursued by us, the spin-charge
gauge symmetry is treated in a sort of MFT while the slave
particle symmetry is kept exact. An opposite choice has been
adopted by Weng (Weng {\it et al} 2000, Weng 2003), {\it i.e.} the
charge and the (reduced) spin symmetries are kept exact as much as
possible, but the slave-particle symmetry is treated in MFT.

 The second major choice in the MF approaches is the
spinon order parameter: if AM-like {\it i.e.}, $<\sum_\alpha
s_{\alpha i}^*  s_{\beta j} >$ (Affleck and Marston 1988) and
nonvanishing, it leaves unbroken the global slave-particle
symmetry but breaks the global spin $SU(2)$; if RVB -like {\it
i.e.} $<\sum_{\alpha \beta}\epsilon_{\alpha\beta} s_{\alpha i}
s_{\beta j}>$  (Wen and Lee 1996, Lee {\it et al} 1998) and
non--vanishing, it breaks the global slave particle symmetry but
preserves the global spin $SU(2)$.

 In the standard slave--boson approach to PG one has
$<$RVB$> \not =0$, though in the $SU(2)$--slave boson approach a
$U(1)$ subgroup is unbroken. A kind of slave-boson approach where
in PG one has$ <$AM$>\not =0$ is pursued in Feng {\it et al}
(2004).

 The choice of the MF order parameter has relevant
physical consequences, because the Ioffe and Larkin (1989) rule
does not appear naturally compatible with a MF in which the
slave-particle symmetry is broken.

 Since in the spin-charge gauge approach we have $<$AM$>
\not  = 0$, but $<$RVB$>$ =0 in MF  in both PG and SM 'phases' the
Ioffe-Larkin rule holds. Furthermore since in MF $|<$AM$>|$=1, the
semion approach seems to optimize both the $t$ and the $J$ terms,
as discussed in 1D (but here a rigorous proof is lacking) in
compatibility with an AF structure.

\section{Spin-charge gauge approach: the key ingredients}

\nin     To summarize, the key physical ingredients of the
spin-charge gauge approach  of MFT to characterize  the `normal
state' within the $t-J$ model are: the Gutzwiller projection,
tackled with spin-charge decomposition, the two-dimensionality
needed for semion statistics, finally leading to a spinon gap, and
the bipartite lattice structure needed  for both AF and  the
$\pi-0$ flux crossover, which is also a peculiar 2D phenomenon.

\subsection{Energy optimization and $\pi-0$ flux crossover}

\nin    Here we focus on the  $\pi-0$ flux crossover.  A key step
is a `semiclassical' treatment of the interaction of $\vec B$ in
the holon hopping term $t h_i^* h_j U_{<ij>}$, where $U_{<ij>}$is
a complex abelian gauge field given by

\be \label{12} U_{<ij>} = e^{i \int^j_i \vec B (y) \cdot d\vec y}
s^*_{\alpha i} (Pe^{i \int^j_i \vec V (y) \cdot d\vec
y})_{\alpha\beta} s_{\beta j}. \ee

\noindent If $p$ denotes a plaquette and $<i j> \in \partial p$
the links in the boundary of $p$, the argument of
$\prod_{<ij>\in\partial p} U_{<ij>}$ plays the role of a magnetic
flux.

 It has been rigorously proved by Lieb (1994) that at
half-filling ($\delta =0)$ the optimal configuration for a
magnetic field on a square lattice in 2D has a flux $\pi$ per
plaquette at arbitrary temperature. On the other hand, it is well
known that at low densities and high temperatures the optimal
configuration has $0$ flux per plaquette.
 At $T=0$ it has been proven that the ground state energy
has a minimum corresponding to one flux quantum per spinless
fermion (Bellisard and Rammal 1990). Numerical simulations (Qin
2000) suggest that increasing $T$ gives rise to a competition
between these minima and at sufficiently high $T$ only $0$ and
$\pi$ flux survive. We expect that the perturbation introduced by
the $J$ term in the $t-J$ model changes the boundary, but not the
essence of the phenomenon of the $\pi \longrightarrow 0$ crossover
itself, while further enhancing the `disorder' lowers the
temperature where only $\pi$ and $0$ flux states survive.

 In the spin-charge gauge MFT it is assumed that the
region of $\pi$ flux of $U$ corresponds to PG and that of $0$-flux
to SM, thus  the `melting of the $\pi$-flux lattice' underlines
the physics of crossover between the two `phases', which in
particular produces a modification in the structure of the
FS.\footnote{Let us remark that even in principle not all
phenomena usually referred to as `pseudogap' should fit in this
interpretation, but only those referring to the $T^*$ temperature
discussed below.}

The optimization considerations discussed above suggest a MF
approximation for $\vec B$ given by \be \label{13} \vec B (y)
\longrightarrow \vec B_{MF} (y)= {1\over 2} \sum_j \vec\nabla \arg
(y - j), \ee

\noindent which gives flux $\pi$ per plaquette. Furthermore, to
obtain the $\pi \longrightarrow 0$ crossover in the MF treatment
of the holon hopping one imposes, by a suitable spinon
redefinition, that
\begin{equation} \nonumber
\prod_{<ij> \in\partial p} (s_{\alpha i}^* (Pe^{i\int^j_i \vec
V(y)\cdot d\vec y})_{\alpha\beta} s_{\beta j} )\sim
   \left\{
   \begin{array}{ll} \label{14}
 1  & {\rm PG} \\
      e^{i \pi}
      & {\rm SM}
   \end{array}
   \right .
\nonumber \end{equation}
 \noindent so that the effect of $\vec
B_{MF}$ on holons in SM is cancelled by the spinon contribution.
The unbalanced presence of $\vec B_{MF}$ in PG produces the
Hofstadter phenomenon: it converts the spinless holon $h$ of SM
with dispersion

\be \label{15} \omega \sim 2 t [(\cos k_x + \cos k_y) - \delta ]
\ee
\noindent and Fermi momenta $k_F \sim 1 - \delta$, {\it i.e.}
a `large FS' roughly consistent with band calculation, into a pair
of `Dirac fields', with pseudospin index corresponding to the two
N\'eel sublattices and dispersion relation:

\be \label{16} \omega \sim 2t (\sqrt{\cos^2 k_x + \cos^2 k_y} -
\delta) \ee \noindent restricted to the magnetic Brillouin zone,
thus yielding a `small FS' centered at ($\pm {\pi\over 2}, \pm
{\pi\over 2})$ with Fermi momenta $k_F \sim \delta$.  As we will
see, many crossover phenomena from PG to SM in transport physics
are due to this change of holon Fermi momenta.

\subsection{Generation of the spinon mass}

\nin Comparing \eref{13} and \eref{9} one can see that the MF
approximation for $\vec B$ consists in neglecting  fluctuations of
$h$; it is therefore reasonable in a MF treatment of $\vec V$ to
neglect the spinon fluctuations {\it i.e.}
 \be \label{17} \vec
V (y) \longrightarrow \vec V_{MF}^a (y) = \delta^{a3} \sum_j
(-1)^{|j|} \vec\nabla \arg (y-j) (1-h^*_j h_j). \ee

\noindent The first term of $\vec V^a_{MF}$ can be gauged away ,
while the second term describes spin vortices, centered at the
holon positions, of opposite vorticity (or chirality) for holons
on two N\'eel sublattices. These chiral spin vortices are
reminiscent of those introduced by Shraiman and Siggia (1988) in
analyzing the single hole problem, but here their density is
finite $\delta> 0$.

  Due to their AF nature the spinons in this approach have
a low-energy effective action described by a $O(3)$ non-linear
$\sigma$-model plus an interaction with  spin vortices appearing
in $\vec V_{MF}$. Due to vortex chirality, the spatial average of
the interaction term linear in $\vec V_{MF}$ vanishes, hence in MF
approximation the leading spinon--vortices interaction term is
given at large scales by $ < \vec V_{MF}^3 \cdot \vec V_{MF}^3>
s_\alpha^* s_\alpha $ where $< >$  denotes spatial average
\footnote{Self-consistently the AF LRO  is assumed absent}.
Naively one expects this average to be proportional to $\delta$,
as this is the density of holons, hence of vortices. However, due
to the long--range tail of vortices one actually gets ( Marchetti
{\it et al} 1998) a logarithmic correction, which, we will see
later on,  is physically relevant. Therefore, as a result of the
interaction with vortices, in MF the spinons acquire a mass gap
$m_s$ given by

\be \label{18} m^2_s \sim <\vec V^3_{MF} \cdot \vec V^3_{MF} >
\sim \delta |\ln \delta |. \ee

\noindent The gap generation here is a kind of wave localization
phenomenon: The gapless waves (here spinons) interacting with a
finite (supercritical) density of impurities (here the vortices),
are getting localized, acquiring a gap. The AF massive structure
of the spinon in our approach prevents boson condensation at low
$T$, in contrast to the slave--boson approach.

  Since the spinon gap has an obvious relation with the AF
correlation length, $\xi_{AF}$, its appearance implies that the
finite (supercritical) density of vortices convert the long-range
AF of the undoped ($\delta =0)$ material into a short--range AF
and the derived doping dependence (roughly $\xi_{MF} \sim
\delta^{-1/2}$) is consistent with the experimental data in
neutron experiment in LSCO (Keimer {\it et al } 1992). Actually,
by including the spinon thermal mass $m_T$ of the renormalized
classical treatment of the $O(3)$--model both the order of
magnitude and the qualitative $\delta - T$ dependence of the
neutron data (Keimer {\it et al } 1992) are reproduced. The $T$
independent term found in experiments which is hard to explain in
terms of a standard $O(3)$ model, as in the spin--fermion approach
(Abanov {\it et al} 2003),  is here exactly due to the
introduction of the spin vortices, a unique feature of this
approach.

\subsection{Reizer singularity of the slave particle gauge field}

\noindent As already remarked, in slave--particle approaches a
slave--particle $U(1)$ gauge field couples spinon and holon. In
the spin-charge gauge approach this can be identified with the
gauge field, $A_\mu$, of the $O(3)$-model rewritten in  the $CP^1$
from, {\it i.e.} the Lagrangian for spinon is
 \be \label{19} L (A,s) \sim {1 \over
J} \Bigl[|(\partial_\mu - A_\mu) s_\alpha |^2 + m^2_s s^*_\alpha
s_\alpha \Bigr] \ee
 \noindent with $\mu=0,1,2$. It turns out that
$A_\mu \sim \tilde s_\alpha^* \partial_\mu \tilde s_\alpha$,
therefore it is staggered.

 The same gauge field appears in the covariant
derivative, obtained via Peierls substitution, acting on holons.
Integrating out spinons and holons one obtains the low--energy
effective action for $A_\mu$. Since $s_\alpha$ is AF and gapful,
the spinon contribution is Maxwellian with a thermal mass $m_0$
for $A_0$. On the contrary,  holons have a finite FS (in both  PG
and SM) and their contribution to the transverse component
exhibits the so--called Reizer singularity (Reizer 1989a,b).

 As a result the leading behavior of the A propagator is:
with $i,j =1,2$ \be \label{20} < A^T_i A_j^T > (\omega,q) \sim
(\delta_{ij} - {q_i q_j\over \vec q^2}) (i \kappa {\omega\over
|\vec q|} - \chi \vec q^2)^{-1}, \ee

\noindent where $\kappa \sim k_F$ is the Landau damping and $\chi
= \chi_s + \chi_h$ with $\chi_s \sim {J\over 6 \pi m_s}, \chi_h =
{t\over 6 \pi k_F} $, is the total diamagnetic susceptibility,
dominated, at least for small $\delta$, by the holon contribution.

 There is a characteristic scale of gauge fluctuations
emerging from the Reizer singularity at finite $T$. It is a kind
of anomalous skin depth, derived by assuming as typical energy
$\omega \sim T$, with the consequence that the transverse gauge
interaction is peaked at $q= Q_T = ({\kappa T\over \chi})^{1/3}$.

 The Reizer singularity in this approach plays a crucial role
in the interpretation of transport properties, and we remark that
it is due to the simultaneous appearance of a finite FS and a gap
for the bosonic excitations $s_\alpha$; if these bosons condense
it disappears. This is what happens at low temperatures in the
slave boson approach, which thus has the difficulty that the
Ioffe-Larkin rule cannot be extended below the holon condensation
temperature.

  The scalar component $A_0$ has a low energy propagator given
by

\be \label{21} < A_0 A_0 > (\omega, q) \sim (\kappa (1+ i
{\omega\over |\vec q|})H(|\vec q|-|\omega|) +m^2_0)^{-1}. \ \ee

  In view of the constant term in \eref{21} the
interaction mediated by $A_0$ is short ranged, hence subleading at
large distance w.r.t. the interaction mediated by $A^T$, triggered
by the Reizer (1989a,b) singularity.

\section{The effect of gauge fluctuations on correlation functions}

\nin A difficulty in  considering the gauge fluctuations is the
lack of a `small parameter' for  expansion. Also, a perturbative
treatment would be insufficient to get a bound state or a
resonance with the electron quantum numbers out of a spinon and a
holon, whereas some kind of binding is expected on the basis of
 the considerations leading to the Ioffe-Larkin rule. A possible way out is
to implement the idea of binding using an eikonal approach in
studying the hydrogen atom as a bound state of a proton and
electron. To analyze the behavior of correlation functions of
physical, hence gauge invariant, fields we apply a kind of eikonal
resummation of (transverse) gauge fluctuations (Marchetti {\it et
al} 2004a). This resummation is obtained by treating first $A_\mu$
as an external field, expanding the correlation function in terms
of first-quantization Feynman paths, then integrating out $A_\mu$
to obtain an interaction between paths which is then treated in
the eikonal approximation. Finally a Fourier transform is
performed to get the retarded correlation function. Further
approximations are
 needed however, especially in the treatment of short-scales, to
get the final result; we refer the reader to Marchetti {\it et al}
(2004a, 2005) for details, but we briefly comment on some
interesting features encountered in the calculation of, for
example, the magnon correlation function.

\subsection{Magnon correlation function}

\nin It turns out numerically that for $\delta, T$ small,
(identified with PG) the spatial Fourier transform in the eikonal
approximation is dominated by a nontrivial complex $|\vec x|$-
saddle point, $|x|_{s.p.} \sim Q_T^{-1}  e^{i \pi/4}$ due to the
effect of gauge fluctuations. The self-consistency requirement for
this eikonal approximation yields a region of validity given
approximately by $ m_s Q_T \lesssim T/\chi \lesssim m_s^2  $
(inserting numbers as order of magnitude from a few tens to a few
hundred K). The upper bound temperature, $T^*$, roughly coincides
both as order of magnitude and as $\delta - T$ dependence (for low
doping) with the pseudogap temperature, slowly decreasing with
$\delta$ due to a delicate cancellation, where the relevance of
the logarithmic correction in \eref{18} shows up: $T^* \sim \chi
m_s^2 \sim {t \over 6 \pi\delta}|\delta\ln \delta| \sim {t \over 6
\pi} |\ln \delta|$. The lower bound temperature $T_{sg}$, with a
maximum at $\delta \sim 0.02$ and then slowly decreasing,  is
reminiscent of the spin-glass crossover in the heavily underdoped
region.

The main effect of the complex saddle point within the above range
is to induce a shift in the mass of spinons: $m_s \rightarrow
M=(m_s^2- i c {T / \chi})^{1/2}$, where $c \sim 3$ is a constant,
thus introducing a dissipation proportional to $T$.  Physically,
 due to the Ioffe-Larkin rule electron conductivity is dominated
by spinons, and the competition between the mass gap and the
dissipation appearing in $M$ is responsible for the MIC of
in-plane resistivity and of many other crossovers in PG, as
discussed already in Sec. 2.

For sufficiently high $\delta $ or $T$ (region identified with SM)
one can verify numerically that the saddle point contribution is
negligible w.r.t. the contribution of fluctuations around 0 in the
range $|\vec x| \lesssim Q_T^{-1}$. The result of gauge fluctuations
can also  be summarized in terms of the appearance of a dissipative
term: $m_s \rightarrow m_s -i c' {T Q_T / \chi m_s^2}$, with $c'
\sim 0.1$.  In SM, however, the change  of $k_F$ from $\delta$ to
$1- \delta$ yields a decrease of diamagnetic susceptibility $\chi$,
implying that the thermal de Broglie wave length is shorter than its
PG counterpart. Therefore the spin-gap effects
($\xi_{AF}<\lambda_T$) are less effective, being confined to very
low temperatures. At a pictorial level the existence of a
characteristic scale  $|\vec x| \sim Q_T^{-1}$ in PG seems to
indicate that spinon and antispinon in the magnon have this typical
size, while  in SM $ Q_T^{-1}$ is just the typical range of their
spatial fluctuations.
 Our approximate time-Fourier transform involves an UV cutoff $\lambda
  Q_T^{-1}$ in terms of a parameter $\lambda$ fixed phenomenologically
   by comparison with experimental data. It turns out that $\lambda <<
   1$ in PG and in SM the result is self-consistent only if $m_s Q_T^{-1},
    {t \over \chi m_s} \lesssim 1 $ with $\lambda \simeq 1$ (numerically
    yielding a range  from a few tens to a few hundred K ). The
retarded magnon propagator derived in this manner is given by:
\begin{equation}
 \langle \vec \Omega \cdot \vec \Omega \rangle (\omega, \vec q)
       \sim
   \left\{
   \begin{array}{ll} \label{27}
       Z_\Omega^{PG}\frac {1}{\omega -2 M}
   J_0( |\vec q|  (2Q_T)^{-1} e^{i \pi/4})   & {\rm PG} \\
     Z_\Omega^{SM}\frac {1} {\omega-2m_s+i\frac{T}{\chi m_s^2} c'
Q_T\lambda}  e^{- \frac {|\vec q|^2}{a}} \frac {1}{a}& {\rm SM}
   \end{array}
   \right .
\end{equation}
\nin where $J_0$ is the Bessel function, $a \sim \frac{T}{\chi
m_s}{Q_T}$ and $\vec q$ is measured from the AF wave vector.

 For the wave function renormalization constant we have $Z_\Omega^{PG}
  \sim Q_T^{-1} (k_F M)^{1/2}$ with $k_F \sim \delta$ and $Z_\Omega^{SM}
   \sim Q_T^{-2} k_F m_s$ with $k_F \sim 1-\delta$.
Equation \eref{27} implies that, within the approximation scheme
adopted, the transverse gauge fluctuations couple the
spinon-antispinon pair in $\vec \Omega$ into a magnon resonance
with mass gap

\be
  m_\Omega \sim 2 m_s
\ee
 \nin and inverse life time
\begin{equation}
  \Gamma_\Omega  =
   \left\{
   \begin{array}{ll} \label{28}
      \Im M \sim T \ {\rm if} \ m_s^2 >> {c T \over \chi},
      \sim T^{1/2} \ {\rm if} \ m_s^2 \sim {c T \over \chi}
      & {\rm PG}  \\
      \frac{T}{\chi m_s^2} c'
Q_T\lambda \sim T^{4/3}
      &{\rm SM}
   \end{array}
   \right  .
\end{equation}
\nin As briefly discussed below, $\Gamma_\Omega^{-1}$ is
proportional to to the electron life-time, which thus always
increases as $T$ decreases. How then this is compatible with MIC?
The point is that above MIC the increase of the electron life-time
yields an increase of conductivity, but below MIC the mobility
decreases as $T$ lowers because the spinon can move only through
thermal duffusion, induced by gauge fluctuations due to gapless
holons, and, because of binding, the electron can move only if the
spinon does. Finally we expect that the effect of the neglected
residual short-range attraction mediated by $A_0$ gives a
renormalization of the mass gap,  without introducing
significative qualitative changes.

\subsection{Electron Propagator}

\nin Let us now briefly comment on the electron correlation
function. Previously we discussed the effect of gauge fluctuations
on the $\Omega$ correlation function at large scales, using the
first quantization path-integral representation for the massive
spinon propagator. An analogous representation is hard to use for
the holon correlation function because of the finite density of
holons. This representation would in fact contain a series of
alternating sign contributions, corresponding to an arbitrary
number of closed fermion wordlines, describing the contributions
of $h$-particles in the finite-density ground state (see {\it
e.g.} Fr\"ohlich and Marchetti 1992). To overcome this difficulty,
we apply a dimensional reduction by means of the tomographic
decomposition introduced by Luther (1978) and Haldane (1994). To
treat the low-energy degrees of freedom we choose a slice of
thickness $k_{F}/\Lambda$, with $\Lambda \gg 1$, in momentum space
around the FS of the holon. We decompose the slice in
approximately square sectors; each sector corresponds to a
quasi-particle field in the sense of Gallavotti-Shankar
renormalization (Benfatto and Gallvotti 1990, Shankar 1994) (see
also Fr\"ohlich 1994). Each sector is characterized by a unit
vector $\vec{n}(\theta)$, pointing from the center of the FS to
the center of the box, labelled by the angle $\theta$ between this
direction and the $k_{x}$ axis. The contribution of each sector
can be viewed approximately as arising from a quasi 1D chiral
fermion; this avoids the finite FS problem for the path-integral
representation discussed above. The paths appearing for a sector
are straight lines directed along the Fermi momenta of the sector,
with small gaussian transverse fluctuations. We now multiply this
path representation for the holon by that for the spinon,
following the decomposition formula \eref{8} for the electron
treated in MF. Finally we integrate out in the eikonal
approximation the transverse gauge fluctuations, proceeding as for
the magnon correlation function. The output for the retarded
correlation function becomes simple only for momenta on the FS,
however the main features presented here hold in a slab of scale
$Q_T$ around the FS. On the FS the result can be summarized as
follows: \be \label{29}
   G^R(\omega,\vec k_F) \sim \frac{Z}{\omega +i\Gamma}.
\ee

In this approximation the scattering rate $\Gamma$ comes entirely
from the spinon, hence it is half of $\Gamma_\Omega$. The
wave-function renormalization constant comes as a product of a
spinon and a holon contribution plus the effect of the quasi 1D
structure of holons yielding for (the isotropic component of) $Z$:
\be \label{30}
   Z \simeq (Q_T m_s)^{1/2}.
\ee
 \nin By chiral invariance, physical absence of superconducting
gap, the mass term induced by the spinon should be exactly
cancelled by the holon-spinon short range attraction. Notice that
the same assumption underlines the recovery of the FS in the
$SU(2)$ slave-boson approach (Lee {\it et al} 2006). We should
emphasize that the electron FS , coinciding in our treatment with
the holon FS, in PG is the `small FS' centered at $(\pm {\pi \over
2}, \pm {\pi \over 2})$ with $k_F \simeq \delta$ whereas in SM it
is the `large FS' with $k_F \simeq 1- \delta$.

In the PG phase  besides \eref{30} we have an additional  angular
dependent factor  around $e.g.$  $({\pi \over 2},{\pi \over 2})$
given by ${1 \over 2}[1 - {1 \over \sqrt{2}} [\cos(\arg \vec k_F)+
\sin( \arg \vec k_F)]]$ of the wave function renormalization
yielding a reduction of the spectral weight outside the magnetic
Brillouin zone, reminiscent of the `Fermi arcs' of ARPES in
underdoped cuprates ( see Wen and Lee (1996), Lee {\it et al}
(1998) for a similar situation in the $SU(2) \times U(1)$ slave
boson approach). This anisotropy is due to the matrix structure of
the `Dirac' hamiltonian, describing physically the
`zitterbewegung', and the relation between the original spinless
holon and the two species of `Dirac holons' introduced by
$B_{MF}$, which changes in the four quarters of the Brillouin zone
centered at $(\pm {\pi \over 2}, \pm {\pi \over 2})$, thus
recovering the standard translational invariance for the electron.
We conjecture (Marchetti {\it et al} 2007) that an additional
reduction of the spectral weight appears as a consequence of the
attraction between spin vortices of opposite vorticity, neglected
in the present MF, improving the agreement with ARPES data.

The structure exhibited by \eref{29} shows that the gauge
fluctuations are able to bind together spinon and holon into a
resonance for low  energies at the Fermi momenta, but with a
strongly temperature dependent wave function renormalization
constant. In particular $Z \sim T^{1/6}$, so  $Z$ vanishes if
formally extrapolated to $T=0$. This implies a peculiar non-Fermi
liquid character for this system of `electron resonances'. In
fact, by gauge invariance the sum of spinon and holon currents is
zero, but at $T=0$ the spinons cannot move as they are gapped and
without any possibility of thermal diffusion, therefore the
electron system behaves as an insulator. A similar phenomenon is
discussed within Weng's approach in Kou and Weng (2005). Existence
of a FS is then compatible only because $Z=0$. However, a real
extrapolation to $T=0$ cannot be done because of the range of
validity of the approximation, as discussed above. The system
therefore appears to fit naturally within the scheme of unstable
fixed points (UFP) outlined by Anderson (2002).

\section{Transport properties}

\nin In this section we sketch the results of calculation for
physical transport quantities through Kubo formulas using the
above approach and compare the theoretical results with the
experimental data.

\subsection{In-plane resistivity}

\nin The in-plane resistivity $\rho$ is calculated via the Ioffe
and Larkin (1989) addition rule: $\rho=\rho_s + \rho_h$, where
$\rho_s$ is the resistivity of the spinon-gauge subsystem and
$\rho_h$ of the holon-gauge subsystem, which is subdominant and of
standard form for a gauge-fermion system (Nagaosa and Lee 1990,
Lee and Nagaosa 1992). Via the Kubo formula:
\begin{eqnarray} \label{kubo}
   (\rho_s)^{-1} & =
    2 \int_0^\infty dx^0 x^0\langle j^{s} j^{s} \rangle (x^0, \vec
    q=0),
\end{eqnarray}
\nin where  $\vec j^s \sim \partial \vec\Omega \sim Q_T
\vec\Omega$ is the spinon current  and, using \eref{27} one
obtains in PG:
\begin{equation}
  \rho \sim \rho_s\sim {|M|^{1/2} \over \sin({\arg M \over 2})} \sim
   \left\{
   \begin{array}{ll}\label{rpg}
      T^{-1}
      & m_s^2 >> {c T \over \chi}  \\
      T^{1 \over 4}
      &m_s^2 \sim {c T \over \chi}
   \end{array}
   \right.
\end{equation}
\nin From \eref{rpg} one recovers the MIC and, due to the square
root in $M$, an inflection point $T^* \sim \chi m_s^2$ at higher
temperature, found also experimentally (Takagi {\it et al} 1992).
Hence, in our approach the MIC is due to correlation effects, not
to a disorder-induced localization, as discussed in Sec. 2. In
fact, within the range of validity for deriving (\ref{rpg}) (above
the N\'eel temperature and the `spin-glass' temperature)  the
insulating behavior is power-law like, not exponential. We believe
that outside this range the qualitative behavior should be the
same, but the formula itself cannot be applied directly.

Moreover, the normalized resistivity
$\rho_n=(\rho-\rho(T_{MIC}))/(\rho(T^*)-\rho(T_{MIC}))$ is a
function only of the ratio $y= cT/T^*$ : $\rho_n(y) = (1 +
y^2)^{1/8} / \sin( {1 \over 4} \arctan y)$. A universal behavior
of this quantity in fact can be verified in experimental data and
in general terms has already been empirically noted in
publications (Wuyts {\it et al} 1996, Konstantinovic {\it et al}
2000). Furthermore, the dependence of $|M|$ on $T^2$ yields an
approximate linear behavior of in-plane resistivity in $T^2$ , for
temperatures just above
 the minimum in PG (Marchetti {\it et al} 2006), observed experimentally (Ando {\it et al}
2004), without involving `Landau' quasi-particles advocated in
(Ando {\it et al} 2004).

The introduction of a magnetic field perpendicular to the plane
allows one to find MIC also for higher dopings suppressing
superconductivity, and it yields a shift of the MIC to higher
temperature which in turn produces a big positive transverse
magnetoresistance (Marchetti {\it et al} 2001), as found in
experiments in LSCO (Lacerda {\it et al} 1994, Kimura {\it et al}
1996, Abe {\it et al} 1999).

 Non-magnetic impurities simply affect the resistivity by an upward
 shift as they appear via the Matthiesen rule only in $\rho_h$, consistent
  with irradiation experiments (Rullier-Albenque {\it
et al} 2003). In contrast, Zn doping  provides strong potential
scattering changing the holon resistivity, and at the same time
disturbs the AF background, making the AF correlation shorter,
therefore qualitatively we expect a shift of MIC in spinon
resistivity up in $T$, as observed in Zn-doped YBCO (Segawa and
Ando 1999). This strong contrast of behavior for non-magnetic
(affecting the holon contribution to  resistivity) and magnetic
(affecting the spinon contribution) scattering, is  indirect proof
for the Ioffe-Larkin rule: {\it It is the resistivity, not the
conductivity of the constituent slave particles} which is added to
form the resistivity of the physical system.

On the other hand, in SM one obtains :
\begin{equation}\label{asympt}
\rho_s\simeq 2 \lambda Q_T^{-1} (\Gamma +  m_s^2/ \Gamma) =
\frac{c'T}{\chi m_s^2} \lambda^2 + \frac{4 m_s^4 \chi}{c' T
Q_T^2},
\end{equation}
\nin where $\lambda \approx 0.7$. In the high temperature limit
$Q_T \gg m_s$, the damping rate in (\ref{asympt}) dominates over
the spin gap $2m_s$ and the spinon contribution to resistivity is
linear in $T$, with a slope $\alpha \simeq (1-\delta)/(\delta |\ln
\delta|)$.
 Lowering the temperature, the second term in
(\ref{asympt}) gives rise first to a superlinear behavior and
then, at the margin of validity of our approach, an unphysical
upturn. The deviation from linearity is due to the spin gap
effects and is cut off in the underdoped samples by the crossover
to the PG phase. We expect that physically in the overdoped
samples it is cut off  by a crossover to a FL`phase'.

The celebrated (approximate) `$T$-linearity' can also be
understood qualitatively as a consequence of $\Gamma$ and the
`effectiveness' of the gauge fluctuations to form resonance,
predominantly contributing to conductivity (see previous Section),
in a slab of momenta $Q_T \sim T^{1/3}$ around the Fermi surface.
In fact the conductivity derived from the Boltzmann transport
theory would be $\sigma_0 \sim \Gamma ^{-1}$, but due to
`effectiveness' the physical conductivity is $\sigma \sim \sigma_0
Q_T \sim T^{-4/3} T^{1/3} \sim T^{-1}$ . Similar considerations
apply to the linearity found in SM for out-of-plane resistivity,
spin relaxation time ${}^{63}(TT_1)$ and, replacing $T$ by
$\omega$ of the AC conductivity (Marchetti {\it et al} 2005), as
discussed below.

\subsection{In-plane IR electronic AC conductivity}

\nin We compute the electronic AC conductivity  in the two limits
$\omega << T$, $T <<
  \omega$, where $\omega$ is the external frequency.
  It turns out that up to the logarithmic accuracy one can pass from the first
  to the second limit by replacing  $T$ with $\omega$ in $Q_T$ and $\Gamma$
  ( we denote the
  obtained quantities by $Q_\omega, \Gamma_\omega$) and rescaling $\Gamma$
   by a positive multiplicative factor
  $\tilde\lambda \lesssim 1/2$.
In PG, in the range $T \lesssim \omega \lesssim J m_s$ the AC
conductivity can be approximately obtained from $\rho^{-1}$
derived from eq. \eref{rpg} substituting $T$ with $\tilde\lambda
\omega$. Hence it exhibits a broad maximum at a frequency
$\omega_{MIC}$ (self-consistently $<< J m_s$), corresponding to a
temperature slightly higher than $T_{MIC}$, as in fact
experimentally seen in Dumm {\it et al} (2003). An $a-b$
anisotropy of $T_{MIC}$ and $\omega_{MIC}$ found in underdoped
LSCO (Dumm {\it et al} 2003) would be generated naturally in the
above scheme by an $a-b$ anisotropy of the magnetic correlation
length (Marchetti {\it et al} 2004b). The anisotropy of $T_{MIC}$
strengthens our interpretation of correlation-induced MIC, since a
disorder-induced localization is expected to have a unique MIC
temperature. As $T$ becomes greater than $\omega$ the AC
conductivity becomes approximately $\omega$ independent.

In SM, for   $\omega << T$ we have
\begin{equation}
\sigma(\omega,T) \sim \frac{Q_T}{i (\omega -2 m_s) +\Gamma} \sim
\frac{1}{i (\omega -2 m_s)
 T^{-1/3} + T},
\end{equation}
\nin while for $T << \omega$
\begin{equation}
\sigma(\omega,T) \sim \frac{Q_\omega}{i (\omega -2 m_s)+
\tilde\lambda \Gamma_\omega}
 \sim \frac{1}{i (\omega -2 m_s) \omega^{-1/3} + \omega}.
\end{equation}
 From the above formulae the following features found experimentally
 in overdoped LSCO (Startseva {\it
et al} 1999) and BSCO (Lupi {\it et al} 2000)
 are easily derived: besides the standard tail $\sim \omega^{-1}$,
 the effect of replacing  $Q_T$ by $Q_\omega$
 is to asymmetrize the peak at $2 m_s$, appearing in Re$\sigma(\omega,T)$
 for $\omega << T$
 and to shift it towards lower frequency.

\subsection{ Spin-lattice relaxation rate ${}^{63}(1/T_1T)$}

\nin Assuming $Q_T$ as the cutoff for the $|\vec q|$ integration
and using the smoothness of the hyperfine field at this scale we
derive in PG:
\begin{eqnarray} \nonumber
 &{}^{63}({1 \over T_1T})  \sim (1-\delta)^2 |M|^{-\frac{1}{2}}
 (a\cos(\frac{\arg M}{2})+ b\sin(\frac{\arg M}{2})) \sim
   \left\{
   \begin{array}{ll}
     a + b T
      & m_s^2 >> {c T \over \chi}  \\
       T^{-{1 \over 4}}
      &m_s^2 \sim {c T \over \chi}
   \end{array},
   \right.
\end{eqnarray}
\nin with $a/b \sim 0.1,$ and the two terms are due to Re and Im
of $J_0$ in \eref{27}.

Therefore we obtain a broad peak as observed in some cuprates
(Berthier {\it et al} 1994) and up to a multiplicative constant a
universality curve as a funcion of $y=c T/T^*$ can be derived.

 In
SM one finds
\begin{equation}
\label{tt163} {}^{63}(\frac{1}{T_1T}) \sim (1-\delta)^2
\rho_s^{-1}.
\end{equation}

Therefore, in the high temperature limit we recover the linear in
$T$ behavior for
 ${}^{63}(T_1T)$ and at higher dopings /lower temperatures
 the superlinear deviation,  also found experimentally in overdoped
 samples of LSCO (Berthier {\it
et al} 1997, Fujiyama {\it et al} 1997).

\subsection{Out-of-plane resistivity}

\nin We assume an incoherent transport along c-direction. $\rho_c$
thus is dominated by virtual hopping between layers and it is
calculated via the Kumar-Jayannavar formula  (Kumar and Jayannavar
1992, Kumar {\it et al} 1997, Kumar {\it et al} 1998):
\begin{equation} \label{kuma}
   \rho_{c} \sim \frac{1}{\nu}\left(\frac{1}
   {\Gamma}+\frac{\Gamma}{t_c^2 Z^2}\right),
\end{equation}
\nin  where  $t_c$ in our  approximation is an effective average
hopping in the $c$-direction
 and $\nu$ the density of states at the Fermi energy.
In PG the first term in \eref{kuma} dominates, yielding an
insulating behavior:
\begin{equation} \nonumber
\rho_c\sim (|M| \sin(\arg M))^{-1}
  \sim
   \left\{
   \begin{array}{ll}
      T^{-1}
      & m_s^2 >> {c T \over \chi}  \\
       T^{-{1 \over 2}}
      &m_s^2 \sim {c T \over \chi}
   \end{array}
   \right .
\nonumber \end{equation}

\nin This crossover reproduces a knee experimentally found in Ito
{\it et al} (1991) and  Yan {\it et al} (1995). A very strong
decrease with $T$ of the anisotropy ratio $\rho_c (T)/\rho_{ab}
(T)$ in PG is also well reproduced by the above formulae.

In the SM phase the second, metallic, term in \eref{kuma}
dominates and substituting $\Gamma$ and $Z$ one recovers the
$T-$linearity in the `incoherent regime' $\Gamma \gg t_c Z$:
\begin{equation}\label{rc}
\rho_c \simeq \frac{J^2}{t_c^2 m_s \nu}\frac{T}{\chi m_s^2}\simeq
\frac{T}{(\delta|\ln\delta|)^{3/2}}. \end{equation}

\nin A universal behavior is also recovered for underdoped samples
in the form $\rho_c=\alpha_c(\delta) \rho_c(y)$ with $\rho_c(y)
\sim (1 + y)^{-1/4} / \sin ({1 \over 2} \arctan y)$ in PG, $\sim
y$ in SM and the $\delta$ dependence of $\alpha_c$ is in agreement
with the phenomenological proposal of universality  (Su {\it et
al} 2006).

\vspace{1pc}

\section{Concluding Remarks}

\nin   As mentioned in the Introduction, there is no consensus
theory of HTS cuprates yet. What has been briefly summarized in
this paper is a theoretical attempt to describe the
metal-insulator crossover as the most spectacular phenomenon of
cuprates in the pseudogap phase, as well as  other transport
properties of cuprates in the pseudogap and `strange metal' phases
in a self-consistent way, using the spin-charge gauge approach. To
the best of our knowledge, this is the only available theoretical
treatment of this crossover as an intrinsic property of the doped
Mott insulators. We start from a two-dimensional $t-J$ model in
the limit $t\gg J$ to describe the strong correlation effects and
use the gauge field approach to implement the single-occupancy
constraints. In the mean field theory we have chosen, the spin
degrees of freedom are represented by hard core bosons, while the
charge degrees of freedom are represented by spinless fermions,
with a slave particle gauge field  binding them `loosely'.

    We have compared the calculated results with experiments,
and in comparing the temperature-doping dependence of the in-plane
resistivity, including the metal-insulator crossover (Marchetti
{\it et al}, 2000, 2001, 2004a) there are no adjustable
parameters, except for the absolute scale of resistance. Moreover,
the nontrivial doping/magnetic field dependence of the
metal-insulator crossover temperature is a direct consequence of
the theoretically predicted spinon mass gap $m_s^2\sim
\delta|\log\delta|$, where $\delta$ is the doping concentration,
and the competition of the short-range antiferromagnetic order
with dissipative motion of the charge carriers. Using the
metal-insulator crossover as a clue to understanding the nature of
the pseudogap phase, we could explain a number of peculiar
transport properties in the pseudogap phase including the
`universality' and anisotropy of in-plane conductivity, the huge
positive magnetoresistance in underdoped samples, the `knee' in
$c$-axis resistivity, the broad peak in the nuclear relaxation
rate and low-frequency spin fluctuation spectrum, within the same
framework, using the same set of parameters.

Using the `melting' of the $\pi$ flux lattice as the main
signature of crossover from the pseudogap to the strange metal
phase, we could easily generalize our approach to higher
doping/temperature and calculate various transport properties to
compare with experimental data in this phase. Apart from
recovering the well-known linear temperature dependence of
in-plane and out-of-plane resistivity, NMR relaxation time,
$\omega^{-1}$ asymptotic behavior of the AC conductivity,  {\it
etc}, we could also account for some  more subtle effects like the
deviation from linearity and the asymmetry of the AC conductivity
peak (Marchetti {\it et al} 2005).

Within the gauge field approach it is natural to consider the
low-energy excitations as `composite particles' consisting of
slave particles loosely bound by the gauge field. A question
arises: Is this gauge field  a pure mathematical construction or
does it reflect to some extent the physical reality? In our
opinion, the second is true. In strongly correlated systems the
spin and charge degrees of freedom are neither `completely
separated' nor `bound in a point-like particle', so the gauge
field is an `indispensable companion' of slave particles
describing the charge and spin degrees of freedom. It took some
time for condensed matter physicists to accept the quasi-particle
as a real physical entity. The quasi-particle is a constituent of
collective motion, and it ceases to exist as soon as it is taken
away from the condensed matter. In a similar way, the gauge field
disappears as soon as the strong correlation is switched off. In
this sense, it is reasonable to consider low-energy excitations in
strongly correlated systems as composite particles bound by the
gauge field.

We would like to sincerely thank J.H. Dai, L. De Leo, G. Orso and F.
Ye for their collaboration at different stages of this project.

\section{References}

\begin{itemize}

\item[]Abanov Ar,  Chubukov A V, and  Schmalian J 2003 {\it Adv.
Phys.} {\bf 52} 119 ({\it Preprint} cond-mat/0107421)

\item[]Abe Y {\it et al} 1999 {\it Phys. Rev.} B {\bf 59} 14753

\item[] Affleck I and  Marston J B  1988 {\it Phys. Rev.} B {\bf
37} 3774

\item[]Anderson P W 1973 {\it Mater. Res. Bull.} {\bf 8} 153

\item[] Anderson P W 1987 {\it Science} {\bf 235} 1196

\item[]Anderson P W 2002 {\it Physica} B {\bf 318} 28

\item[] Anderson P W, Lee P A, Randeria M, Rice T M, Trivedi N,
and Zhang F C 2004 {\it J. Phys.: Condens. Matter} {\bf 16} R755

\item[] Ando Y {\it et al} 1995  {\it Phys. Rev. Lett.} {\bf 75}
4662

\item[]Ando Y {\it et al} 1996a   {\it Phys. Rev. Lett.} {\bf 77}
2065

\item[] Ando Y  {\it et al} 1996b {\it J. Low Temp. Phys.} {\bf
105} 867

\item[] Ando Y {\it et al} 1999 {\it Phys. Rev. Lett.} {\bf 83}
2813

\item[] Ando Y {\it et al} 2001 {\it Phys. Rev. Lett.} {\bf 87 }
017001

\item[]Ando Y {\it et al} 2004 {\it Phys. Rev. Lett.} {\bf 92}
197001

\item[] Arovas D P and Auerbach A 1988 {\it Phys. Rev.} B {\bf 38}
316

\item[] Baskaran G and Anderson P W 1988 {\it Phys. Rev.} B {\bf
37} 580

 \item[]Baskaran G, Zou Z, and Anaderson P W 1987 {\it
Solid State Commun.} {\bf 63} 973

\item[]Bednorz J G and M\"uller K A 1986 {\it Z. Phys. B: Condens.
Matter} {\bf 64} 189

\item[]Bellisard J and  Rammal R 1990 {\it Europhys. Lett.}   {\bf
13} 205

\item[]Benfatto G and  Gallavotti G 1990  {\it J. Stat. Phys.}
{\bf 59} 541

\item[]Berthier C {\it et al} 1994  {\it Physica} C {\bf 235-240}
67

\item[]Berthier C {\it et al} 1997  {\it J. Physique}  I {\bf 6},
2205

 \item[]Boebinger G S {\it et al} 1996 {\it Phys. Rev. Lett.} {\bf
77} 5417

\item[] Chen W and  Ng Y J 1995 {\it Phys. Rev.} B {\bf 51}, 14479
(1995)

\item[] Damascelli A, Hussain Z, and  Shen Z X 2003 {\it Rev. Mod.
Phys.} {\bf 75} 473

\item[] Dorey N and  Mavromatos N E 1991 {\it Phys. Rev.} B {\bf
44} 5286

\item[] Dumm M, Basov D N , Komiya S, and  Ando Y 2003 {\it Phys.
Rev. Lett.} {\bf 91} 077004

\item[]Farakos K  and  Mavromatos N E 1998 {\it Phys. Rev.} B {\bf
57} 3017

\item[]Fazekas P and Anderson P W 1974 {\it Philos. Mag.} {\bf 30}
432

 \item[]Feng S P, Qin J H , Ma T X 2004 {\it J. Phys.: Condens.
Matter} {\bf 16} 343

\item[] Fiory A T {\it et al} 1990 {\it Phys. Rev.} B {\bf 41}
 2627

\item[]Fournier P {\it et al} 1998 {\it Phys. Rev. Lett.} {\bf 81}
4720

\item[]Fradkin E 1991  {\it Field Theories of Condensed Matter
Systems} (Addison-Wesley - New York)

\item[]Fr\"{o}hlich  J and Marchetti P A (1992) {\it Phys. Rev.} B
{\bf 46} 6535

\item[]Fr\"ohlich J {\it et al} 1994   {\it Fluctuating Geometries
in Statistical Mechanics and Field Theory, Proc.  Les Houches
1994}

\item[]Fujiyama S {\it et al} 1997  {\it J. Phys. Soc. Japan.}
{\bf 66}, 2864

\item[] Gros C 1988 {\it Phys. Rev.} B {\bf 38} 931

\item[] Gros C 1989 {\it Ann. Phys.} (N.Y.) {\bf 189} 53

\item[]Ha Z N C and Haldane F D M 1994 {\it Phys. Rev. Lett.} {\bf
73} 2887

\item[]Haldane F D M 1991 {\it Phys. Rev. Lett.} {\bf 67} 937

\item[]Haldane F D M  1994  {\it Perspectives in Many-Particle Physics,
Proc. Int. School  Phys. `Enrico Fermi', Course CXXI }, ed. R A
Broglia  and J R Schrieffer (Amsterdam North-Holland), p.5

\item[]Ioffe L and Larkin A 1989 {\it Phys. Rev.} B {\bf 39} 8988

\item[]Ito T {\it et al} 1991 {\it Nature} (London) {\bf 350} 596

\item[]Keimer B {\it et al} 1992 {\it Phys. Rev.} B {\bf 46} 14034

\item[]Kimura T {\it et al} 1996 {\it Phys. Rev.} B {\bf 53} 8733

\item[]Kivelson S A, Rokhsar D S, and Sethna J P 1987 {\it Phys.
Rev.} B {\bf 35} 8865

\item[]Konstantinovic Z {\it et al} 2000 Physica C {\bf 341} 859

\item[]Kotliar G and Liu J 1988 {\it Phys. Rev. } B {\bf 38} 5142

\item[]Kou S P and Weng Z Y 2005 {\it Eur. Phys. J.} B {\bf 47} 37

\item[]Kumar N and  Jayannavar A M 1992 {\it Phys. Rev.}  B {\bf
45} 5001

\item[]Kumar N {\it et al} 1997  {\it Mod. Phys. Lett.} B {\bf 11}
347

\item[]Kumar N {\it et al} 1998 {\it Phys. Rev.} B {\bf 57} 13399

\item[]Lacerda A {\it et al} 1994 {\it Phys. Rev.} B {\bf 49} 9097

\item[]Laughlin R B 1988 {\it Science} {\bf 242} 525

\item[]Laughlin R B 1997 {\it Phys. Rev. Lett.} {\bf 79} 1726

\item[]Lee P A and Nagaosa N 1992 {\it Phys. Rev.} B {\bf 46} 5621

\item[]Lee P A, Nagaosa N, Ng T K, and Wen X G 1998 {\it Phys.
Rev.} B {\bf 57} 6003

\item[] Lee P A, Nagaosa N and Wen X G 2006 {\it Rev. Mod. Phys.}
{\bf 78} 17

 \item[]Lieb E H 1994 {\it Phys. Rev. Lett.}
{\bf 73} 2158

 \item[]Lupi S {\it et al} 2000 {\it Phys. Rev.} B {\bf 62}
12418

\item[]Luther A 1979 {\it Phys. Rev.} B {\bf 19} 320

\item[]Marchetti P A,  De Leo L, Orso G, Su Z B and Yu L
2004a {\it Phys. Rev.} B {\bf 69} 024527

\item[]Marchetti P A , Orso G,  Su Z B and Yu L 2004b {\it Phys.
Rev.} B {\bf 69} 214514

\item[]Marchetti P A, Orso G, Su Z B  and Yu L 2005 {\it Phys.
Rev.} B {\bf 71} 134510

\item[]Marchetti P A, Su Z B, Yu L 1996  {\it Nucl. Phys.} B {\bf
482} 731

\item[]Marchetti P A, Su Z B and Yu L 1998 {\it Phys. Rev.} B {\bf
58}  5808

\item[]Marchetti P A, Su Z B and  Yu L 2000 {\it J. Phys.: Condens
Matt} {\bf 12} L329

\item[]Marchetti P A, Su Z B and  Yu L 2001 {\it Phys. Rev. Lett.}
{\bf 86} 3831

\item[]Marchetti P A , Su Z B, Yu L 2006 {\it Proc.  8th
Intern. Conf. on Materials and Mechanisms of Superconductivity and
High Temperature Superconductors}, 9-14 July, Dresden, to appear in
{\it Physica C}

\item[]Marchetti P A , Su Z B, Yu L  2007 in preparation

\item[]Nagaosa N and  Lee P A 1990 {\it Phys. Rev. Lett.} {\bf 64}
2450

\item[] Norman M and Pepin C 2003 {\it Rep. Prog. Phys.} {\bf 66}
1547

\item[]Ogata M and Shiba H 1990 {\it Phys. Rev.} B {\bf 41} 2326

\item[]Ono S and Ando Y 2003 {\it Phys. Rev.} B {\bf 67} 104512

\item[]Ono S {\it et al} 2000 {\it Phys. Rev. Lett.} {\bf 85} 638

\item[]Onose Y {\it et al} 2001 {\it Phys. Rev. Lett.}
 {\bf 87} 217001

\item[] Orgad D {\it et al} 2001 {\it Phys. Rev. Lett.} {\bf 86}
4362

\item[] Paramekanti A, Randeria M, and Trivedi N 2001 {\it Phys.
Rev. Lett.} {\bf 87} 217002

\item[]Paramekanti A, Randeria M, and Trivedi N 2003 {\it
Preprint} cond-mat/0305611

\item[]Qin G 2000, unpublished

\item[]Reizer M 1989 {\it Phys. Rev.} B {\bf 39} 1602; {\bf 40}
11571

\item[]Rullier-Albenque F, Alloul H, and  Tourbot R 2003 {\it
Phys. Rev. Lett.} {\bf 91} 047001

\item[]Segawa K and Ando Y 1999  {\it Phys. Rev.} B {\bf 59} R3948

\item[]Shankar R 1994  {\it Rev. Mod. Phys.} {\bf 66} 129

\item[]Shraiman B I, and Siggia E D  1988 {\it Phys. Rev. Lett.}
{\bf 61} 467

\item[]Startseva  T {\it et al} 1999 {\it Physica} C {\bf 321} 135

\item[]Su Y H,  Luo H G, and  Xiang T 2006  {\it Phys. Rev.} B
{\bf 73} 134510

\item[]Takagi H {\it et al} 1992 {\it Phys. Rev. Lett.} {\bf 69}
2975

\item[]Timusk T and Statt B 1999 {\it Rep. Prog. Phys.} {\bf 62}
61

\item[] Trappeniers L {\it et al} 1999 {\it J. Low Temp. Phys.}
{\bf 117} 681

\item[]Wen X G and Lee P A 1996 {\it Phys. Rev. Lett.} {\bf 76}
503

\item[]Weng Z Y 2003 {\it Proc. of  Int. Symp. on Frontiers of Science
in honour of C N Yang, Beijing }(Singapore: World Scientific),
{\it Preprint} cond-mat/0304261, and references therein

\item[]Weng Z Y, Sheng D N  and Ting C S 2000 {\it Physica} C {\bf
341-348} 67

\item[] Wu Y S 1994 {\it Phys. Rev. Lett.} {\bf 73} 922

\item[]Wuyts B {\it et al} 1996 {\it Phys. Rev.} B {\bf 53} 9418

\item[]Yan Y F {\it et al} 1995  {\it Phys. Rev.} B {\bf 52} R751

\item[]Yoshioka S 1989  {\it J. Phys. Soc. Japan.}  {\bf 58} 32

\item[] Zhang F C, Gros C, Rice T M, and Shiba H 1988 {\it
Supercond. Sci. Technol.} {\bf 1} 36

\item[]Zou Z and Anderson P W 1988 {\it Phys. Rev.} B {\bf 37} 627

\end{itemize}

\end{document}